\documentclass{kluwer}    % Specifies the document style.

\newdisplay{guess}{Conjecture}
\usepackage{psfig}

\begin{document}                                                                                   
\begin{article}
\begin{opening}         
\title{Present Evidence for Intermediate Mass Black Holes in ULXs and
Future Prospects} 
\author{J. M. \surname{Miller}}  
\runningauthor{J. M. Miller}
\runningtitle{IMBHS and ULXs}
\institute{Harvard-Smithsonian CfA, 60 Garden Street, Cambridge MA
02138, USA; jmmiller@cfa.harvard.edu}
%\date{Oct 18 2004}

\begin{abstract}
In a number of the most luminous ULXs (those with $L_{X} \sim
10^{40}~{\rm erg}~{\rm s}^{-1}$) in nearby galaxies, observations with
{\it XMM-Newton} and {\it Chandra} are revealing evidence which
suggests that these ULXs may harbor intermediate-mass black holes
(IMBHs).  The detection of accretion disk spectral components with
temperatures 5--10 times lower than the temperatures observed in
stellar-mass black hole binaries near to their Eddington limit may be
particularly compelling evidence for IMBH primaries, since $T \propto
M^{-1/4}$ for disks around black holes.  In some sources, X-ray timing
diagnostics also hint at IMBHs.  Evidence for IMBHs in a subset of the
most luminous ULXs, a discussion of the robustness of this evidence
and alternatives to the IMBH interpretation, and prospects for better
determining the nature of these sources in the future, are presented
in this work.

\end{abstract}
%\keywords{sample, \LaTeX}

\end{opening}           

\section{Introduction to ULXs}  

Ultraluminous X-ray sources (ULXs) are bright, off-nuclear point
sources in nearby normal galaxies, for which the inferred X-ray
luminosity exceeds the isotropic Eddington limit for a $10~M_{\odot}$
black hole ($L \simeq 1.3 \times 10^{39}~ {\rm erg}~ {\rm s}^{-1}$;
Frank, King, \& Raine 2003).  In most ULXs, long-timescale X-ray
variability indicates that the ULXs are likely accreting black hole
binaries like those known in the Milky Way.  The existence of ULXs was
first revealed with {\it Einstein} (Fabbiano 1989).  {\it Chandra} and
{\it XMM-Newton} have revolutionized the study of these sources by
enabling observations which effectively isolate individual sources and
which obtain sensitive spectra and lightcurves from the brightest
nearby examples.  For a recent review of ULXs, see, e.g., Fabbiano \&
White (2005).

Among the reasons that ULXs are interesting is that they may represent
rare phases of accretion in binary systems, rare X-rays states, and/or
rare phases of binary evolution.  However, the most compelling reason
to study ULXs is likely because their luminosity suggests that they
may harbor intermediate mass black holes (IMBHs, $10^{2-5}~M_{\odot}$;
for a recent review see Miller \& Colbert 2004).  It is well-known
that stellar-mass black holes can reach luminosities slightly in
excess of their implied isotropic Eddington limit (see, e.g.,
McClintock \& Remillard 2005), so ULXs at the lower end of the
luminosity range (indeed, this is the majority of ULXs) are likely
binaries with stellar-mass black hole primaries (or neutron star
primaries in rare cases).  However, those ULXs at the upper-end of the
luminosity range (arbitrarily, $L_{X} \geq 10^{40}~ {\rm erg}~ {\rm s}^{-1}$) may harbor IMBHs.  Recent
observations with {\it Chandra} and {\it XMM-Newton} (in particular)
have revealed cool accretion disk components in the spectra of the
most luminous ULXs (see Miller, Fabian, \& Miller 2004a) and other
X-ray spectral and timing features which suggest IMBHs.  The status of
the evidence for IMBHs in ULXs, arguments against IMBHs, and prospects
for better understanding the IMBH candidates is discussed below.

\begin{figure}[t] 
\begin{center} 
\hbox{
\psfig{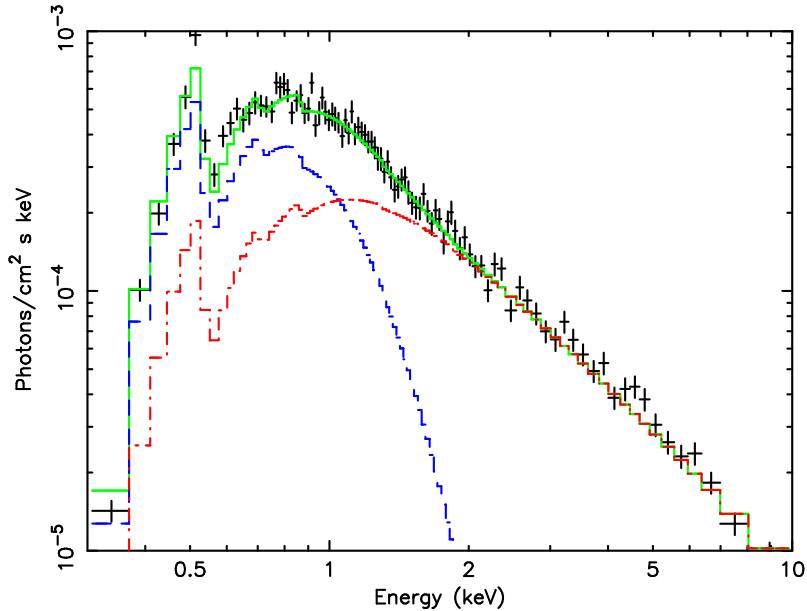}
%\hspace{0.1cm}
%\psfig{figure=lca.ps,width=0.52\textwidth,angle=-90}
%\hspace{1.0cm}
}
\vspace{-1.0cm} 
\end{center} 
\caption{\footnotesize {The {\it XMM-Newton}/EPIC-pn X-ray spectrum of NGC 1313 X-1 is shown
above (Miller, Fabian, \& Miller 2004b).  A disk component is shown in
blue, and hard power-law emission is shown in red.  In this source and
a growing number of other ULXs, X-ray spectra require a cool disk component,
which may be evidence for an accreting IMBH, e.g. in a binary system.} }
\label{fig1} 
\end{figure}

\begin{figure}[t] 
\begin{center} 
\hbox{
\psfig{figure=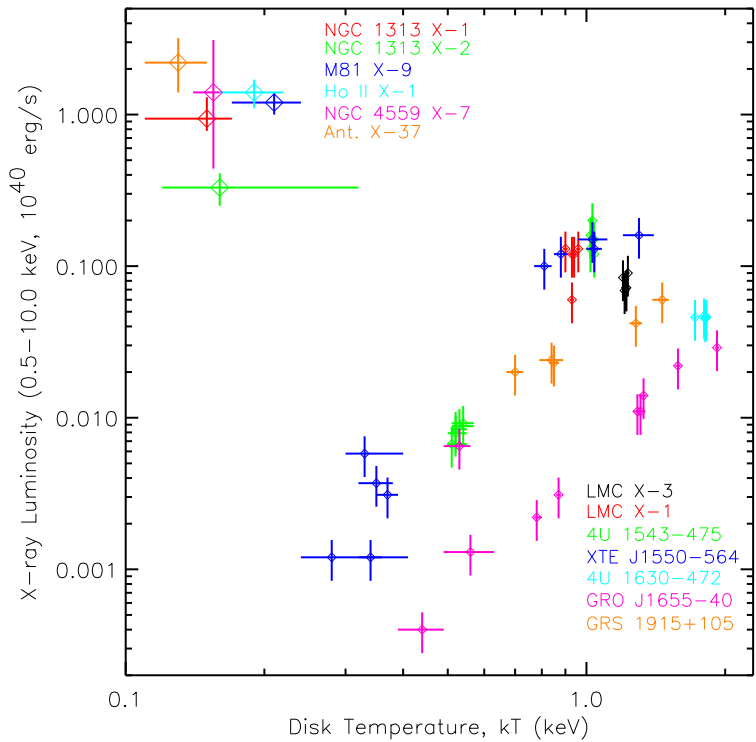,width=0.9\textwidth}
%\hspace{0.1cm}
%\psfig{figure=lca.ps,width=0.52\textwidth,angle=-90}
%\hspace{1.0cm}
}
\vspace{-1.0cm} 
\end{center} 
\caption{\footnotesize {ULXs with cool accretion disks
do not lie on the temperature-luminosity trend observed in
stellar-mass black holes, and form a rather tight group, suggesting a
distinct sub-class which may indeed harbor IMBHs (Miller, Fabian, \&
Miller 2004a).} }
\label{fig1} 
\end{figure}

\begin{figure}[t] 
\begin{center} 
\hbox{
\psfig{figure=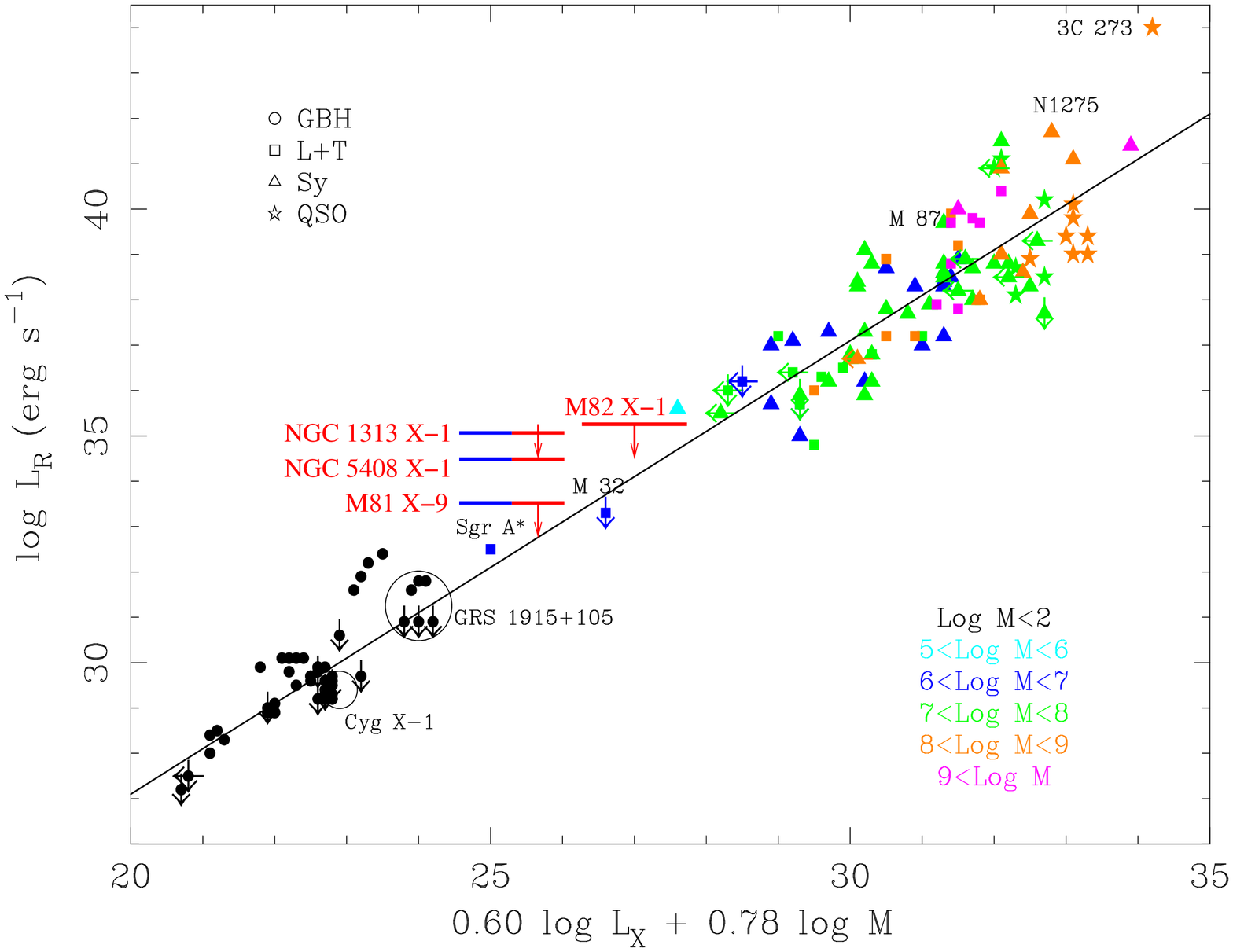,width=0.9\textwidth}
%\hspace{0.1cm}
%\psfig{figure=lca.ps,width=0.52\textwidth,angle=-90}
%\hspace{1.0cm}
}
\vspace{-1.0cm} 
\end{center} 
\caption{\footnotesize {Merloni, Heinz, \& Di~Matteo (2003) found a
``fundamental plane'' relating black hole mass, X-ray luminosity, and
radio luminosity in unbeamed sources.  Bright ULXs with cool disks and
good limits on radio flux (NGC 1313 X-1: Colbert et al.\ 1995; M81
X-9: $F\leq 60~ \mu Jy$ at 5~GHz, Jimenez-Garate et al. 2005 in prep.)
or radio detections (NGC 5408 X-1: Kaaret et al.\ 2003), and M82 X-1,
fit between stellar-mass and supermassive black holes on the
fundamental plane.  For M82 X-1, the radio limits and lower mass bound
are from Kaaret et al.\ (2001), and the upper mass bound is from
Strohmayer \& Mushotzky (2003).  For the other sources, red error bars
represent a mass range of $300-3000~M_{\odot}$, a representative range
for disk normalization and temperature scaling, respectively, and the
blue error bars extend down to approximately the mass found from
simple Eddington limit scaling.  (The fundamental plane makes use of
2--10~keV luminosity values and flux densities at 5~GHz.  ULX
luminosities reported in Miller, Fabian,
\& Miller 2004a were converted to the 2--10~keV band and flat radio
spectra were assumed). } }
\label{fig1} 
\end{figure}

\section{Evidence for IMBHs in ULXs}

Spectra of bright ULXs obtained with {\it ASCA} were typically
described in terms single flux components, and often in terms of hot
accretion disk components (Makishima et al.\ 2000), although weak
evidence for two-component spectra consisting of cool accretion disk
and hard power-law flux components was also reported (Colbert \&
Mushotzky 1999).  Recent observations of the most luminous and
proximal ULXs with {\it XMM-Newton}~ and {\it Chandra} appear to have
resolved this discrepancy: Miller et al.\ (2003) first reported the
detection of a cool accretion disk at the 8$\sigma$ level of
confidence in {\it XMM-Newton} spectra of NGC~1313 X-1 (see Figure 1).
Other significant detections of cool disk components have been
reported in {\it XMM-Newton}~ and/or {\it Chandra} spectra of NGC 1313 X-2
(Miller et al.\ 2003), NGC 5408 X-1 (Kaaret et al.\ 2003), M81 X-9
(Holmberg IX X-1, Miller, Fabian, \& Miller 2004a), Antennae X-37
(Miller et al.\ 2004), Holmberg II X-1 (Dewangan et al.\ 2004), NGC
4559 X-7 (Cropper et al.\ 2000), and most recently in a ULX in M101
(Kong, DiStefano, \& Yuan 2004).

In each of these cases, the luminosity of the ULX is near to or above
$L_{X} \simeq 10^{40}~ {\rm erg}~ {\rm s}^{-1}$, and the measured disk
color temperature is consistent with the $kT \simeq 0.1-0.2$~keV
range.  These ULX disk temperatures are much lower than the $kT \simeq
1-2$~keV disks temperatures commonly measured in $\sim 10~M_{\odot}$
black hole binaries in the Milky Way and LMC when they are observed at
luminosities near to their Eddington limit.  Indeed, because $T
\propto M^{-1/4}$ for standard disks around black holes (Frank, King,
\& Raine 2002), the low disk temperatures measured in these ULXs imply
black holes with masses in the $few \times 10^{2-3}~M_{\odot}$ range
when scaled to the temperatures seen in stellar-mass black hole
binaries.  (The normalizations of these disk components can also be
scaled and generally imply IMBHs; see Miller et al.\ 2003.)  These
ULXs occupy a distinct region of a $L_{X} - kT$ diagram, which is
distinct from the $L \propto T^{4}$ trend clearly seen in stellar-mass
black holes (see Figure 2; see also Miller, Fabian, \& Miller 2004a).
The exceptionally high luminosities inferred in these sources, their
low disk temperatures, and their clustering in a $L_{X} - kT$ diagram
makes these sources strong IMBH-candidates.

M82 X-1 may be the single best IMBH-candidate ULX presently known, in
part due to its exceptionally high luminosity, which approaches $L_{X}
\simeq 10^{41}~ {\rm erg}~ {\rm s}^{-1}$.  However, the nature of its
low energy spectrum is unknown due to contamination from a surrounding
thermal plasma.  Strohmayer \& Mushotzky (2003) have reported the
detection of a 54~mHz QPO in the X-ray flux of this ULX, based on
observations with {\it XMM-Newton} and {\it RXTE}.  As QPOs are
thought to be disk oscillations, this feature indicates that the X-ray
flux is not beamed, and implies the observed luminosity is a true
isotropic luminosity, greatly reinforcing an IMBH interpretation.

A timing diagnostic has been interpreted as tentative evidence for an
IMBH in at least one other case: Cropper et al.\ (2004) have reported
a break frequency at 28~mHz in the power density spectrum of NGC 4559
X-7.  The break could indicate a mass of $38~M_{\odot}$ or
$1300~M_{\odot}$, as there are {\it two} breaks commonly observed in
the power spectra of accreting black holes (both in stellar-mass and
supermassive black holes; see, e.g., Uttley, McHardy, \& Papadakis
2002).  It is very difficult to obtain a mass scaling from either a
single break or a single QPO, and indeed there is still some ambiguity
when two breaks or multiple QPOs are detected since characteristic
frequencies drift with flux.

\section{On the Robustness of Evidence for IMBHs in ULXs}

\subsection{Details of the Spectral Modeling}

Evidence for cool accretion disks in sources with luminosities near to
$10^{40}~ {\rm erg}~ {\rm s}^{-1}$ is generally quite robust.  The
requirement for low disk temperature with normalizations also
suggesting IMBH primaries does not depend the choice of disk model
(Miller, Fabian, \& Miller 2004a).  Cool disks are required regardless
of whether independent flux components or self-consistent Compton
up-scattering models are used to fit the X-ray spectra (Miller et
al. 2003; Miller, Fabian, \& Miller 2004).  An uncertainty in scaling
from disks around stellar-mass black holes to disks around presumed
IMBHs exists due to potential differences between the intrinsic disk
spectrum and measured spectrum (after transfer through a disk
atmosphere).  However, it has recently been shown that these effects
are remarkably similar in disks around stellar-mass black holes and
IMBHs (Fabian, Ross, \& Miller 2004).  Finally, low metal abundances
in absorbing material along the line of sight to these sources does
not falsely create a statistical need for a cool disk component in the
most sensitive spectra (Miller et al. 2003; Miller, Fabian, \& Miller 2004b).

As noted above, spectra of some of the most luminous ULXs obtained
with {\it ASCA} could be described only in terms of a single hot disk
component (temperatures approached 2~keV; see Makishima et al.\ 2000).
It is now clear that these sources are generally better described with
a combination of a cool disk and hard power-law.  Both {\it Chandra}
and {\it XMM-Newton} have much narrower beams than {\it ASCA}, and so
fold-in far less background (both diffuse and point-source in nature).
Moreover, the lower energy bounds of {\it Chandra} and {\it
XMM-Newton} are lower than the effective lower energy bound of {\it
ASCA}, which gradually increased over the mission lifetime.

Stobbart, Roberts, \& Warwick (2004) recently fit the spectrum of a
dipping ULX in NGC 55 ($L_{x} = 1.6 \times 10^{39}~ {\rm erg}~ {\rm
s}^{-1}$) with a model consisting of a hot ($kT=0.8$~keV) disk
dominating the high energy spectrum, and a very soft ($\Gamma = 4$)
power-law dominating the low energy spectrum.  Some parameters were
fixed in fits to the spectra of this ULX (this is not a standard
practice), so it is not clear that a model consisting of a low
temperature disk and hard power-law is inconsistent with the data
(though the spectra of dipping sources can be confusing and
non-standard).  Aside from the fact that the spectral fitting method
is non-standard, the model is inconsistent with the observed spectra
of stellar-mass Galactic black hole binaries and Compton up-scattering
models for hard X-ray emission (Stobbart, Roberts, \& Warwick note
these facts).

Although this model is likely unphysical, it is interesting to test
whether or not it might provide a reasonable alternative (in terms of
a goodness-of-fit statistic) to spectra of IMBH-candidate ULXs which
are presently well described in terms of cool disks and hard power-law
components.  Let us take NGC 1313 X-1 as an example.  Jointly fitting
the {\it XMM-Newton} pn, MOS1, and MOS2 spectra of NGC 1313 X-1 (see
Miller et al.\ 2003; Miller, Fabian, \& Miller 2004) with a simple
model consisting of multicolor disk black body and power-law
components (modified by neutral absorption), the cool disk plus hard
power-law model gives a very good fit: $\chi^{3}/dof = 885.8/870$ 
($kT = 0.18$~keV, $\Gamma
= 1.8$).  Constraining the disk temperature to be $kT
\geq 0.8$~keV, the resultant fit is significantly worse statistically:
$\chi^{2}/dof = 965.3/870$ ($kT = 2.6$~keV, $\Gamma = 4.3$).  

As stated before, M82 X-1 is unlike other IMBH-candidate ULXs in that
it is embedded in a region of strong diffuse plasma emission, which
makes it very difficult to constrain the nature of its low energy
spectrum (see Strohmayer \& Mushotzky 2003).  Although other strong
IMBH-candidate ULXs are not located within similar clusters of young
stars, it is nevertheless important to understand the extent to which
thermal plasmas may contribute to the low-temperature thermal
emission.  At low temperatures, an O~VII or O~VIII emission line
should be particularly strong, and there is no evidence for such a
line in the X-ray spectra of IMBH-candidate ULXs (note that careful
fitting of the O K-edge is required to prevent false O VII/VIII line
detections; see Miller, Fabian, \& Miller 2004a).  At present, no soft
X-ray line has been reported in a ULX which is significant at or above
the $3\sigma$ level.  Thus, it presently appears that the
low-temperature thermal components detected in some ULXs are due to
optically-thick disk components, but improved spectra are required to
put strong constraints on possible contributions from an
optically-thin thermal plasma.

\subsection{Relativistic Beaming}

Relativistic beaming of source flux has been proposed as a means of
explaining luminosities in apparent excess of the isotropic Eddington
limit for a $10~M_{\odot}$ black hole (Reynolds et al.\ 1997;
Kording, Falcke, \& Markoff 2002).  The very low radio to X-ray flux
ratios found in luminous ULXs with cool disks strongly argues against
this interpretation, since beaming tends to create flat $\nu {\rm
F}_{\nu}$ spectra (e.g. in blazars, Miller et al.\ 2003; Kaaret et
al.\ 2003; Miller, Fabian, \& Miller 2004).  Indeed, the ratios found
are below the maximum ratios observed in stellar mass black holes
which are observed {\it edge-on} (see Fender \& Kuulkers 2001), rather
than along a line of sight coincident with a jet axis.  Blackbody
components imply a minimum physical size, and here again the cool
disks detected argue against relativistic beaming.  Finally, it should
be noted that in the special case of M82 X-1, the detection of QPO
strongly argues that the disk is seen clearly, and therefore
argues against relativistic beaming as a viable means of explaining
the inferred luminosity (Strohmayer \& Mushotzky 2003).

It is interesting to consider how the broad-band properties of
IMBH-candidate ULXs with constraining radio limits or detections
compare to stellar-mass and supermassive black holes.  Merloni, Heinz,
\& DiMatteo (2003) examined the properties of a number of unbeamed
black holes, and found a fundamental plane of black hole activity
which relates mass, X-ray luminosity, and radio luminosity.  Figure 3
shows the fundamental plane, with the addition of the IMBH-candidate
ULXs NGC 1313 X-1, M81 X-9, NGC 5408 X-1, and M82 X-1.  These black
holes lie on the fundamental plane, or only slightly above the plane
(but no farther than the scatter in the supermassive black holes
population).  This again implies that these sources are not beamed,
and their position between the stellar-mass black hole population and
supermassive black holes again suggests that they may indeed harbor
IMBHs.

\subsection{Disk Issues: Funnels, Photon Bubbles, and Slim Disks}

It has been suggested that at very high mass accretion rates, a funnel
may form at the inner disk and boost luminosities by factors of 10--30
along the funnel axis (King et al.\ 2001).  This model would allow
$\sim 10~M_{\odot}$ black holes to apparently violate the isotropic
Eddington limit, and could explain the high apparent luminosities seen
in some ULXs.  The low disk temperatures measured in some
IMBH-candidate ULXs already argue against this model, because disk
temperatures in the 0.1--0.2~keV range would only signal a high mass
accretion rate for IMBHs.  Observations of stellar-mass galactic black
holes also argue against the formation of such structures: 4U
1543$-$475 is a Galactic black hole binary viewed at $i = 21^{\circ}$,
but it does not exceed its Eddington limit by more than a factor of a few
(Miller, Fabian, \& Miller 2004b; Park et al.\ 2004).  Stronger
funneling might be required, but radiation will leak out the sides of
a strong funnel geometry as it becomes ionized and optically-thin,
ruining the funnel effect (King \& Pounds 2003).  Finally, in the case
of Holmberg II X-1, optical spectroscopy of a surrounding nebula
reveals that the isotropic luminosity must exceed the Eddington limit
for a $10~M_{\odot}$ black hole (Kaaret, Zezas, \& Ward 2004), and the
morphology of optical nebulae around other ULXs does not strongly
suggest beaming.

It has also been suggested that radiation pressure-dominated disks
might be able to produce super-Eddington fluxes through small-scale
photon bubble instabilities (Begelman 2002).  Slim disk solutions
(e.g. Watarai, Mizuno, \& Mineshige 2001) may also allow high fluxes.
Where IMBH-candidate ULXs are concerned, the difficulty with both
models is that they are expected to hold at very high mass accretion
rates -- when disks are expected to be hot.  The low disk
temperatures observed in some IMBH-candidate ULXs means that
photon-bubble disk and/or slim disks are not required in these sources.

\subsection{Photospheres}

It has recently been suggested that the basic disk plus corona
geometry inferred in black hole systems accreting at high rates may be
incorrect, and that a better model may consist of an outflowing
photosphere which is optically thick at $r \leq 100~R_{Schw.}$ and
external shocks generating hard X-rays (King \& Pounds 2003).  In any
case where the flux in the hard spectral component is equal to the
flux in the soft component, the photosphere would have to flow at $v =
c$ and the photospheric radius would have to equal $1~R_{Schw.}$ to
generate the observed hard X-ray flux, which renders this alternative
implausible in most ULXs with cool disks and luminous quasars.  Weak
X-ray absorption lines in the spectra of some AGN -- apparently
blue-shifted in the frame of the AGN and cited as evidence in favor of
outflowing photospheres -- have recently been shown to coincide with
the AGN recession velocity in a number of cases, demanding absorption
near to the Milky Way (McKernan, Yaqoob,
\& Reynolds 2004).  Thus, both in the case of AGN and ULXs, the data
would seem to argue against this alternative in the majority of cases.

It should be noted that sources like the bright, soft transient in
M101 (Kong, DiStefano, \& Yuan 2004), as well as ``super-soft'' sources
and ``qausi-soft'' sources, do not have significant hard components and
the photosphere model cannot be excluded in these cases.  Apart from
the immediate issue of photospheres, the very low flux observed from
super-soft and quasi-soft sources makes it nearly impossible to obtain
strong spectral and timing constraints, and inferences for IMBHs in
these sources require extreme caution as they are necessarily at the
level of what the data will allow, rather than what the data strictly
require.

\subsection{Inferences from X-ray Luminosity Functions}

It is sometimes argued that X-ray luminosity functions demonstrate
that all ULXs are stellar-mass X-ray binaries, because the highest
luminosity sources appear to extend naturally from the
lower-luminosity distribution without a break (see, e.g., Swartz et
al.\ 2004).  Several assumptions are implicit in such an argument:
first, that IMBHs should have a narrow mass range (a wide range of
masses would act to diminish a break); second, that IMBHs should all
be accreting at high fractions of their Eddington limit (a range of
mass accretion rates like that seen in stellar-mass binaries in the
Milky Way and Magellanic Clouds would also act to flatten a break);
and third, that a single observation or a few observations can reliably
constrain the nature of a source population that is likely to be
variable.

The high luminosity end of the X-ray luminosity function of a given
galaxy is a regime with very few sources, where strong
constraints are not possible.  The statistics do not allow one to
strongly require or to exclude a break.  Consider how different the
X-ray luminosity functions of galaxies like M101 (Kong, DiStefano, \&
Yuan 2004) and NGC 3628 (Strickland et al.\ 2001) must appear when
their transient ULXs (which reach near to or above $L_{X} \simeq
10^{40}~ {\rm erg}~ {\rm s}^{-1}$) vary by factors of 10 and 1000,
respectively.  Again, as this regime is a low-statistics regime, it is
unlikely that one source or even two sources could require a break,
but it illustrates the danger of drawing conclusions about source
populations based on X-ray luminosity functions.  In fact, the
statistical uncertainty of X-ray luminosity functions is worse than
typically presented: errors on luminosity resulting from uncertainties
in the absorbing column and differences between spectral models that
cannot be distinguished in poor quality spectra are not generally
considered.

X-ray luminosity functions do indicate that only a small number of
IMBHs are accreting at high fractions of their Eddington limit at any
given time in a typical galaxy.  This inference is broadly consistent
with the behaviors observed from stellar mass and supermassive black
holes.  Moreover, X-ray luminosity functions may demonstrate that the
population of IMBHs which may exist in binaries is likely small
relative to the number of black holes and neutron stars in accreting
binaries.  These inferences are limited, but as such they are fair to
the limited statistics at the high luminosity end of X-ray luminosity
functions.

\section{Future Prospects}

It will likely prove to be very difficult to obtain optical/IR radial
velocity curves of ULXs to constrain the mass of the primary.  Putting
even the brightest stellar-mass Galactic binaries at distances of a
few Mpc makes them very faint indeed.  Further complications arise
because most IMBH-candidate ULXs -- in contrast to most stellar-mass
Galactic black holes -- are persistently active; light from the
accretion disk will make it difficult to identify and trace features
from the companion star.  If ULXs are very wide binaries, the long
orbital period will make it even harder to obtain radial velocity
curves.  The optical nebulae found around some IMBH-candidate ULXs
(Pakull \& Mirioni 2003) may further complicate spectroscopic studies
of the presumed binary system.  For the foreseeable future, it is
likely that the nature of IMBH-candidate ULXs will be decided based on
a preponderance of indirect evidence.

From an X-ray point of view, much longer observations of galaxies
harboring IMBH-candidate ULXs are urgently needed.  Observations of
300--500~ksec (and longer) have been devoted to the study of
relativistic effects in accreting sources (e.g. 500~ksec to study
putative absorption lines from the surface of the neutron star in
EXO~0748$-$676, nearly 400~ksec to study the broad Fe~K$\alpha$
emission line in MCG--6-30-15, and nearly 300~ksec to study the broad
Fe~K$\alpha$ emission line in GX~339$-$4).  Establishing the presence
or absence of IMBHs -- a new class of relativistic objects -- is a
goal as deserving of long observations as the study of phenomena in
known classes of relativistic objects.

For ULXs within a few Mpc with $L_{X} \geq 10^{40}~ {\rm erg}~ {\rm
s}^{-1}$, an {\it XMM-Newton} observation of 300--500~ksec will
achieve the sensitivity required to: (1) detect a broad Fe~K$\alpha$
emission line from the accretion disk, which would provide independent
evidence for a standard accretion disk and rule-out relativistic and
geometric beaming; (2) detect breaks in the power-density spectrum
and/or QPOs, which will enable independent mass estimates based on
scaling the characteristic frequencies (indeed, the relation between
QPO frequency and break frequency in Galactic black holes and neutron
stars found by Wijnands \& van der Klis 1999 may provide an additional
pragmatic scaling beyond scalings based only on breaks); and (3) detect
any soft X-ray emission lines, enabling the (likely small) flux of any
diffuse optically-thin plasma to be separated from an underlying
optically-thick disk continuum.

Of course, far better radio and optical constraints are also required.
Even though optical/IR radial velocity curves may be difficult to
obtain, the value of spectroscopic studies of the nebulae surrounding
some ULXs is clear (see Pakull \& Mirioni 2003; Kaaret, Ward, \& Zezas
2004).  Independent luminosity and beaming constraints can impact our
view of the nature of IMBH-candidate ULXs.  Better radio constraints
are needed to rule-out beaming in more sources.  Radio detections may
not be possible in some cases, but strong limits allow sources to be
placed on the ``fundamental plane'' (for instance), and the position
of sources on the ``fundamental plane'' does reflect their nature
(though it does not allow for a precise mass measurement).  Mushotzky
(2004) and collaborators have undertaken a radio survey of ULXs, and in some
cases have found broad contours coincident with the X-ray source
positions; the lack of strongly peaked point source emission in most
cases may again argue against beaming.

X-rays probe the regions closest to compact objects, and
can be expected to have the greatest impact on our understanding of
ULXs and IMBH-candidate ULXs in particular.  In the long run, planned
missions like {\it Constellation-X} and {\it XEUS} will revolutionize
the study of ULXs.  These missions will make it possible to obtain
sensitive spectroscopic and timing constraints on ULXs with short
observations.  Relativistic reverberation mapping in AGN may be a
primary goal of missions like {\it Constellation-X} and {\it XEUS},
but for a variety of possible designs, these missions will be able to
reveal or reject the IMBH hypothesis in a much higher number of ULXs.

\section{Summary}

The notorious M82 X-1 and at least six ULXs with luminosities near to
or above $L_{X} \simeq 10^{40}~ {\rm erg}~ {\rm s}^{-1}$ and soft
components consistent with optically-thick emission from cool ($kT =
0.1-0.2$~keV) accretion disks (Miller, Fabian, \& Miller 2004b) may be
regarded as IMBH-candidates.  The properties which have been observed
to separate these ULXs from populations of stellar-mass and supermassive
black holes appear to be robust.  In the cases where they can be
strongly tested, alternative explanations for the high inferred
luminosities and the nature of the spectra appear to be implausible.  This
proceedings has attempted to take a critical view of the evidence
supporting an IMBH explanation for a subset of very luminous ULXs, but
also a critical view of the counter-arguments.  Dramatic strides
towards resolving the nature of these sources can be made in the near
future with optical and radio observations, and principally with very
long X-ray observations.

\acknowledgements
Thanks to A. C. Fabian and M. C. Miller for collaboration and
discussions.  Thanks to A. Merloni for his generous permission to use
the fundamental plane figure, and also to M. Jimenez-Garate for his
generous communication of the radio limit on M81 X-9.  Thanks to
P. Martinez.  Final thanks to the National Science Foundation for
support through their Astronomy \& Astrophysics Postdoctoral
Fellowship.
\theendnotes

\end{article}

\begin{thebibliography}{}

\bibitem[\protect\citeauthoryear{Brown and Burton}{1978}]{BrownAndBurton}
%Ballantyne D., Vaughan S., Fabian A., 2003, MN, 342, 293\\
Begelman, M., 2002, ApJ, 538, L97\\
Colbert, E. J. M., \& Mushotzky, R. F., 1999, ApJ, 519, 89\\
Cropper, M. C., Soria, R., Mushotzky, R., Wu, K., Markwardt, C., \&
Pakull, M., 2004, MNRAS, 349, 39\\
Dewangan, G., Miyaji, T., Griffiths, R. E., \& Lehmann, I., 2004, ApJ,
608, L57\\
Fabian, A. C., Ross, R. R., \& Miller, J. M., 2004, MNRAS, 335, 359\\
Fabbiano, G., 1989, ARA\& A, 27, 87\\
Fabbiano, G., \& White, N. E., 2005, to appear in ``Compact Stellar
X-ray Sources'' eds. M. van der Klis and W. H. G. Lewin, Cambridge
Univ.\ Press, astro-ph/0307077\\
Fender, R. P., \& Kuulkers, E., 2001, MNRAS, 324, 923\\
Frank, J., King, A. R., \& Raine, D., 2002, in ``Accretion Power in
Astrophysics'', Cambridge: Cambridge Univ. Press
Kaaret, P., et al., 2001, MNRAS, 321, L29\\
Kaaret, P., Corbel, S., Prestwich, A. H., \& Zezas, A., 2003, Science,
299, 365\\
Kaaret, P., Ward, M. J., \& Zezas, A., 2004, MNRAS, 351, 83\\
King, A. R., Davies, M. B., Ward, M. J., Fabbiano, G., \& Elvis, M.,
2001, ApJ, 552, L109\\
King, A. R., \& Pounds, K., 2003, MNRAS, 345, 657\\
Kording, E., Falcke, H., \& Markoff, S., 2002, A\& A, 382, L13\\
Kong, A. K. H., DiStefano, R., \& Yuan, F., 2004, ApJ, 617, L49\\
Makishima, K., et al., 2000, ApJ, 535, 632\\
McClintock, J. E., \& Remillard,a R. A., 2005, to appear in ``Compact Stellar
X-ray Sources'' eds. M. van der Klis and W. H. G. Lewin, Cambridge
Univ.\ Press, astro-ph/0306213\\
McKernan, B., Yaqoob, T., \& Reynolds, C. S., 2004, ApJ, subm.,
astro-ph/0408506\\
Miller, J. M., Fabbiano, G., Miller, M. C., \& Fabian, A. C., 2003,
ApJ, 585, L37\\
Miller, J. M., Fabian, A. C., \& Miller, M. C., 2004a, ApJ, 614, L117\\
Miller, J. M., Fabian, A. C., \& Miller, M. C., 2004b, ApJ, 607, 931\\
Miller, J. M., Zezas, A., Fabbiano, G., \& Schweizer, F., 2004, ApJ,
609, 782\\
Miller, M. C., \& Colbert, E. J. M., 2004, IJMPD, 2004, 1\\
Mushotzky, R., 2004, to appear in the proceedings of Kyoto, 2003,
astro-ph/0411040\\
Pakull, M., \& Mirioni, L., 2003, in ``New Visions of the X-ray
Universe in the XMM-Newton and Chandra ERA'' (ESA SP-488; Nordwijk:
ESA), astro-ph/0202488\\
Park S. Q., et al., 2004, ApJ, 610, 378\\
Reynolds, C. S., Loan, A. J., Fabian, A. C., Makishima, K., Brandt,
W. N., \& Mizuno, T., 1997, MNRAS, 286, 349\\
Stobbart, A.-M., Roberts, T. P., \& Warwick, R. S., 2004, MNRAS, 351, 1063\\
Strickland, D., Colbert, E. J. M., Heckman, T. M., Weaver, K. A.,
Swartz, D. A., Ghosh, K. K., Tennant, A. F., \& Wu, K., 2004, ApJS,
154, 519\\
Dahlem, M., \& Stevens, I. R., 2001, ApJ, 560, 707\\
Uttley, P., McHardy, I. M., \& Papadakis, I. E., 2002, MNRAS, 332, 231\\
Watarai, K., Mizuno, T., \& Mineshige, S., 2001, ApJ, 549, L77\\
\end{thebibliography}
\end{document}